\documentclass[12pt]{article}

\usepackage{amssymb}
\usepackage{amsmath}

\usepackage{graphicx}

\begin{document} %
%%%%%%%%%%%%%%%%%%

\newcommand{\bs}{\boldsymbol} % bold symbols in mathematical mode

%%%%%%%%%%%%%%%%%%%%%%%%%%%%%%%%%%%%%%%%%%%%%%%%%%%%%%%%%%%%%%%%%%%%%

\title{Search for invisible dark photon in $\gamma e$ scattering at future
lepton colliders}

\author{
S.C. \.{I}nan\thanks{Electronic address: sceminan@cumhuriyet.tr}
\\
{\small Department of Physics, Sivas Cumhuriyet University, 58140,
Sivas, Turkey}
\\
{\small and}
\\
A.V. Kisselev\thanks{Electronic address:
alexandre.kisselev@ihep.ru} \\
{\small Division of Theoretical Physics, A.A. Logunov Institute for
High Energy Physics,}
\\
{\small NRC ``Kurchatov Institute'', 142281, Protvino, Russia}}

\date{}

\maketitle

\begin{abstract}
For the first time, the production of a massive dark photon (DP) in
the $\gamma e^-$ scattering at the future lepton colliders ILC,
CLIC, and CEPC is examined. The invisible decay mode of the DP is
addressed. We have studied both the unpolarized scattering and the
collision of the Compton backscattered photons with the polarized
electron. The missing energy distributions are shown. We have
considered the wide range 10 GeV -- 1000 GeV of the DP mass
$m_{A'}$. The excluded regions at the 95\% C.L. in the plane
$(\varepsilon, m_{A'})$, where $\varepsilon$ is the kinetic mixing
parameter, are obtained.
\end{abstract}

%%%%%%%%%%%%%%%%%%%%%%%%%%%%%%%%%%%%%%%%%%%%%%%%%%%%%%%%%%%%%%%%%%%%%

%%%%%%%%%%%%%%%%%%%%%%%%
\section{Introduction} %
%%%%%%%%%%%%%%%%%%%%%%%%
\label{sec:intr}

One of the main goals of the present and proposed collider
experiments is to search for dark matter (DM) particles. The DM
makes up approximately 85\% of the total mass in the Universe. Its
existence is firmly confirmed by gravitational experiments
\cite{Trimble:1987, Bertone:2018}, but composition and nature are
still an open question. We consider a DM scenario in which no DM
fields are charged under the SM gauge group, and the lightest stable
DM particles, $\chi$, can only interact with SM through the exchange
of a vector mediator, dark photon (DP) (also known as hidden or
heavy photon) \cite{Fayet:1980}-\cite{Hooper:2012}. It is usually
denoted as $A'$. In its turn, the DP kinetically mixes with the SM
$U(1)_Y$ hypercharge gauge field at the renormalizable level
(\emph{kinematic-mixing portal}) \cite{Holdom:1986_1}. Such mixing
can be generated by loops of massive particles charged under both
$U(1)_Y$ and secluded $U(1)'$ symmetries. That kinetic-mixing portal
model may be extended to 5D by adding one flat ED \cite{Rizzo:2018}.
The existence of a new light dark sector can be also connected to a
generation of neutrino masses \cite{Bertuzzo:2019}. An electroweak
gauge extension of the SM by adding an extra $U(1)'$, with mixing
with the standard $U(1)_Y$ can also result in a new boson
\cite{Aguila:1994}-\cite{Rizzo:1998}. For recent reviews on the DP,
see, for instance, \cite{Raggi:2015}-\cite{Graham:2021}. In
particular, astrophysical constraints on the DP parameters can be
found in \cite{Carlson:1987}-\cite{Bi:2021}, while bounds from
$(g-2)_{\mu,e}$ are presented in
\cite{Bennett:2006,Pospelov:2009_2}.

The main mechanisms of the DP production are meson decays ($\pi^0,
\eta \rightarrow A' \gamma$) \cite{Ilten:2015}, bremsstrahlung ($eZ
\rightarrow eZA'$ and $pZ \rightarrow pZA'$) \cite{Gninenko:2018,
Bjorken:2009}, Drell-Yan ($q \bar{q} \rightarrow A'$), and
annihilation ($e^+e^- \rightarrow A' \gamma$). The process of pair
annihilation into a dark and an ordinary photon provided a striking
benchmark (mono-photon plus missing energy) for the DP search at the
LEP \cite{OPAL:1999}-\cite{DELPHI:2005}. A probing new physics in
final states containing a photon and missing transverse momentum in
proton-proton collisions was presented by the ATLAS and CMS
Collaborations \cite{ATLAS:2014}-\cite{CMS:2019}. For the most
recent results on the DPs from the LHC, see
\cite{ATLAS:2021}-\cite{CMS:2021_2}. The ATLAS and CMS search
sensitivities for DPs at the high luminosity LHC can be found in
\cite{ATLAS:2019_HL,CMS:2018_HL}. A DP phenomenology at the LHC and
HL-LHC was studied in \cite{Curtin:2015}-\cite{Cobal:2020}, and
searches of the DPs at the LHeC and FCC-he colliders were discussed
in \cite{Onofrio:2020}.

A probing new light gauge boson similar to the photon in $e^+e^-$
collisions is of particular interest
\cite{Fayet:2007}-\cite{Chen:2021}. The searches for the DP at
$e^+e^-$ colliders \cite{BaBar:2014}-\cite{Belle II:2018} (see also
review papers \cite{Raggi:2015}-\cite{Graham:2021}) have looked for
its decays to the $e^+e^-, \mu^+\mu^-$ and $\pi^+\pi^-$ final
states, as well as for invisible decays of $A'$. The DP{ production
in the Compton-like $\gamma e \rightarrow A' e$ process is studied
in \cite{Chakrabarty:2019,Smirnov:2021}. The inverse process, $A' e
\rightarrow \gamma e$, is recently considered in
\cite{Hochberg:2021}. The gamma factory's discovery potential
through the low energy dark Compton scattering is analyzed in
\cite{Chakraborti:2021}. In \cite{Wong:2021} the
$\gamma\gamma\rightarrow\gamma A'$ process for MeV scale collider is
considered.

In the present paper we examine a novel DP production based on a
high energy $\gamma e^-$ scattering at the lepton colliders ILC
\cite{ILC_1,ILC_2}, CLIC \cite{CLIC_1,CLIC_2}, and CEPC
\cite{CEPC_1,CEPC_2}, the $\gamma e^- \rightarrow A' e^-$ process.
Both unpolarized and polarized collisions are considered in the wide
DP mass region 10 GeV -- 1000 GeV. We expect the DP to decay
predominantly into \emph{invisible} dark-sector particles. Let us
underline that up to now the DP production at high energy lepton
colliders was studied only for the $e^+e^-$ mode of these colliders
\cite{Blaising1a:2021}-\cite{Kalinowski:2021_conf}.

%%%%%%%%%%%%%%%%%%%%%%%%%%%%%%%
\section{Massive dark photon} %
%%%%%%%%%%%%%%%%%%%%%%%%%%%%%%%
\label{sec:DPs}

As it was already mentioned above, the DP does not directly couple
to SM fields. But there could be a small coupling to the
electromagnetic current $J_\mu$ due to \emph{kinetic mixing} between
the SM hypercharge and the field strength tensor of the DP field.
Consider the case when the dark sector is represented by just a
single extra $U(1)'$ gauge group. Let $B_\mu$, $\bar{A}'_\mu$ be the
mediator fields of the SM $U(1)_Y$ symmetry and dark $U(1)'$ gauge
group. The gauge Lagrangian can be taken in the following form
\begin{equation}\label{Lagrangian_gauge}
\mathcal{L}_{\mathrm{gauge}} = - \frac{1}{4} B_{\mu\nu}B^{\mu\nu} -
\frac{1}{4} \bar{F}'_{\mu\nu}\bar{F}'^{\mu\nu} -
\frac{\varepsilon}{2 c_W} \bar{F}'_{\mu\nu}B^{\mu\nu} \;,
\end{equation}
where $B_{\mu\nu} = \partial_\mu B_\nu - \partial_\nu B_\mu$ and
$\bar{F}'_{\mu\nu} = \partial_\mu \bar{A}'_\nu - \partial_\nu
\bar{A}'_\mu$ are the field strength tensors of $U(1)_Y$ and
$U(1)'$, respectively,  $c_W$ is  the cosine of the Weinberg angle
$\theta_W$, and $\varepsilon \ll 1$ is the kinetic mixing parameter.
The kinetic mixing can be generated at the one-loop level by massive
particles charged under both $U(1)_Y$ and $U(1)'$ symmetries. After
diagonalization of the gauge fields $W^3_\mu$, $B_\mu$, and DP field
$\bar{A}'_\mu$ \cite{Fabbrichesi:2020},
\begin{equation}\label{diagonalization}
\left(
  \begin{array}{c}
    W^3_\mu \\
    B_\mu \\
    \bar{A}'_\mu \\
  \end{array}
\right)
=
\left(
  \begin{array}{ccc}
    c_W & s_W & -s_W \varepsilon \\
    -s_W & c_W & -c_W \varepsilon \\
    t_W \varepsilon & 0 & 1 \\
  \end{array}
\right)
\left(
  \begin{array}{c}
    Z_\mu \\
    A_\mu \\
    A'_\mu \\
  \end{array}
\right) ,
\end{equation}
where $s_W$ and $t_W$ are the sine and tangent of $\theta_W$, we
obtain the interaction Lagrangian up to $\mathrm{O}(\varepsilon^2)$
\begin{equation}\label{Lagrangian_DP}
\mathcal{L}_{\mathrm{int}} = e J_\mu A^\mu- \varepsilon e J_\mu
A'^\mu + \varepsilon e' t_W J'_\mu Z_\mu + e' J'_\mu A'^\mu +
\mathcal{L}_{A'\chi} \;.
\end{equation}
Here $A_\mu$, $Z_\mu$ are the physical gauge fields, and $A'^\mu$ is
the physical field of the DP. $J'_\mu$ and $e'$ are the DM matter
current and DP coupling to the dark-sector matter, respectively. In
\eqref{Lagrangian_DP} we have added the last term which describes a
$A'\chi\chi$ interaction, where $\chi$ is a dark matter particle.
The form of this interaction is left unspecified. As one can see in
\eqref{Lagrangian_DP}, the coupling of the massive DP the SM
fermions is $-\varepsilon e$. The $Z$ gauge boson acquires the
coupling strength $\varepsilon e' t_W$ to the dark sector current.

Thus, there are three unknown parameters: the DP mass, $m_{A'}$, the
mixing parameter, $\varepsilon$, and the
$A'\rightarrow\chi\bar{\chi}$ branching. The latter is taken to be
unity, if we assume that $m_{A'} > 2m_\chi$. The dimensionless
mixing parameter $\varepsilon$ is a priori unknown and presumably
lies in the $10^{-12} - 10^{-2}$ region, depending on the DP mass
\cite{Chang:2018}, \cite{Bjorken:2009},
\cite{Arkani-Hammed:2008}-\cite{Cocoli:2011} (see also Fig.~8.16 in
\cite{Briefing_book}, where sensitivities for the DP in the plane
mixing parameter $\varepsilon$ versus $m_{A'}$ are collected). The
DP mass $m_{A'}$ can also vary in a wide range, but most of
experimental searches for the invisible DP were aimed at the
$m_{A'}$ in the region $1 \mathrm{\ MeV} - 10 \mathrm{\ GeV}$
\cite{Raggi:2015}-\cite{Graham:2021}, see
Fig.~\ref{fig:current_limits_invis}. On the contrary, we will
examine the DP mass region $10 \mathrm{\ GeV} - 1000 \mathrm{\
GeV}$.
%
%%%%%%%%%%%%%%%%%%%%%%%%%%%%%%%%%%%%%%%%%%%%%
% Figure 1. Current limits on DP parameters %
% Invisible decay of DP                     %
%%%%%%%%%%%%%%%%%%%%%%%%%%%%%%%%%%%%%%%%%%%%%
\begin{figure}[htb]
\begin{center}
\includegraphics[scale=0.25]{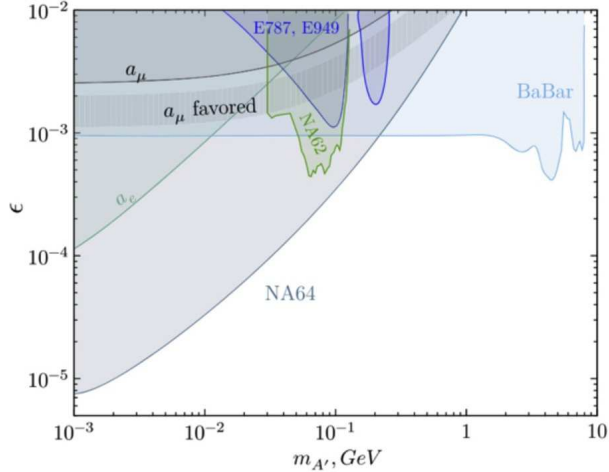}
\caption{The existing limits for the massive dark photon going to
invisible final states. The constraints from $a_e = (g-2)_e$ and
$a_\mu = (g-2)_\mu$ are also shown. The figure is taken from
\protect\cite{Filippi:2020}.} \label{fig:current_limits_invis}
\end{center}
\end{figure}

%%%%%%%%%%%%%%%%%%%%%%%%%%%%%%%%%%%%%%%%%%%%%%%%%%%%%%%%%%%%%%%%%%%%%%%
\section{Invisible dark photon production in Compton-like scattering} %
%%%%%%%%%%%%%%%%%%%%%%%%%%%%%%%%%%%%%%%%%%%%%%%%%%%%%%%%%%%%%%%%%%%%%%%

The production of the DP in the center-of-mass system of the $\gamma
e^-$ scattering,
\begin{equation}\label{process}
\gamma + e^- \rightarrow A' + e^- ,
\end{equation}
is depicted in Fig.~\ref{fig:DP_process}. As is known, linear
$e^+e^-$ colliders can operate in $\gamma e$ and $\gamma \gamma$
modes \cite{Ginzburg:1983}-\cite{Telnov:2016}. The $\gamma\gamma$
facilities at the future circular colliders are examined in
\cite{Aleksan:2015}-\cite{Belusevic:2019}. In particular, a number
of processes in $\gamma e$ and $\gamma\gamma$ collisions at the CEPC
have been studied in \cite{Dev:2018}. At the lepton collider hard
real photons may be generated by the laser Compton backscattering,
when soft laser photons collide with electron beams. As a result, a
large flux of photons is produced which carry a great amount of the
parent electron energy. A $\gamma e$ ($\gamma \gamma$) collider has
a number of advantages over $e^+e^-$ collider. Among they are i)
Higgs can be $s$-channel produced; ii) higher cross sections for
charged particles; iii) higher mass reach in some channels; iv) pure
QED interaction (in $e^+e^-$ a $Z$ boson exchange is present); v)
higher polarization of initial states. The $\gamma e$ and
$\gamma\gamma$ modes  were considered for a number of processes at
the future lepton colliders
\cite{Gou:2017}-\cite{Inan&Kisselev:2021_3}, but the DP production
in $\gamma e$ or $\gamma \gamma$ collisions have not yet been
studied at the ILC, CLIC, or CEPC.
%
%%%%%%%%%%%%%%%%%%%%%%%%%%%%%%%%%%%%%%%%%%%%%
% Figure 2. The process gamma + e -> A' + e %
%%%%%%%%%%%%%%%%%%%%%%%%%%%%%%%%%%%%%%%%%%%%%
\begin{figure}[htb]
\begin{center}
\includegraphics[scale=0.5]{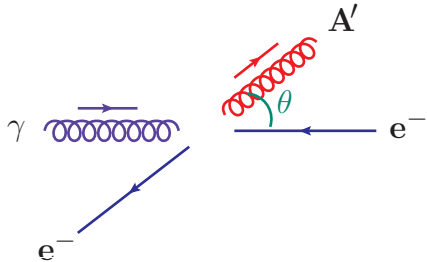}
\caption{The production of the dark photon $A'$ in the $\gamma e^-$
collision.}
\label{fig:DP_process}
\end{center}
\end{figure}

Let $E_0$ be the energy of the initial laser photon beam, $E_e$ be
the energy of the initial electron beam, while $E_\gamma$ be the
energy of the Compton backscattered (CB) photon. The differential
cross section for the unpolarized DP production accompanied by
electron at the lepton collider operating in the $\gamma e$ mode is
defined as
\begin{equation}\label{dif_cross_sec}
\frac{d\sigma}{d\cos\theta} = \int_{x_{\min}}^{x_{\max}} dx
f_{\gamma/e}(x) \,\frac{d\hat{\sigma}}{d\cos\theta} \;.
\end{equation}
Here $f_{\gamma/e}(x)$ is the distribution of the CB photon in the
variable $x = E_\gamma/E_e$,
\begin{equation}\label{x_limits}
x_{\min} = \frac{p_\bot^2}{E_e^2} \;, \quad x_{\max} =
\frac{\zeta}{1 + \zeta} \;, \quad \zeta = \frac{4E_0 E_e}{m^2} \;,
\end{equation}
$p_\bot$ is the transverse momentum of the outgoing particles, $m$
is the electron mass, and $\theta$ represents the scattering angle
of the outgoing DP (see Fig.~\ref{fig:DP_process}). The laser beam
energy $E_0$ is chosen to maximize $E_\gamma$. It is achieved, if
$\zeta = 4.8$, and we get $x_{\max} = 0.83$, see
eq.~\eqref{x_limits}.

The spectrum of the CB photons in formula \eqref{dif_cross_sec} is
defined as follows \cite{Ginzburg:1983}
\begin{equation}\label{photon_spectrum}
f_{\gamma/e}(x) = \frac{1}{g(\zeta)}\left[1-x+\frac{1}{1-x} -
\frac{4x}{\zeta(1-x)} + \frac{4x^{2}}{\zeta^{2}(1-x)^{2}}\right] ,
\end{equation}
where
\begin{equation}\label{g}
g(\zeta) =
\left(1-\frac{4}{\zeta}-\frac{8}{\zeta^2}\right)\ln{(\zeta+1)} +
\frac{1}{2} + \frac{8}{\zeta}-\frac{1}{2(\zeta+1)^2} \;.
\end{equation}
The differential cross section for the process $\gamma e^-
\rightarrow A' e^-$ in the center-of-mass system of the colliding
particles is defined as
\begin{equation}\label{gamma-e_dif_cross_sec}
\frac{d\hat{\sigma}}{d\cos\theta} =\frac{1}{32\pi\hat{s}}
\frac{\sqrt{(\hat{s} + m_{A'}^2 - m^2)^2 - 4\hat{s}
m_{A'}^2}}{\hat{s} - m^2} \,|F(\hat{s}, \cos\theta)|^2 \;,
\end{equation}
where $\sqrt{\hat{s}} = \sqrt{sx}$ is the center-of-mass energy of
the backscattered photon and electron, $\hat{s} \geqslant (m_{A'} +
m)^2$. A matrix element of the process \eqref{process} is a sum of
two diagrams in Fig.~\ref{fig:DP_prod}, in which the $A'e^-e^+$
coupling constant is equal to $-\varepsilon e$. An explicit
expression for the square matrix element $|F|^2$ is given in
Appendix~A.
%
%%%%%%%%%%%%%%%%%%%%%%%%%%%%%%%%%%%%%%%%%%%%%%%%%%%%%%
% Figure 3. Diagrams for process gamma + e -> A' + e %
%%%%%%%%%%%%%%%%%%%%%%%%%%%%%%%%%%%%%%%%%%%%%%%%%%%%%%
\begin{figure}[htb]
\begin{center}
\includegraphics[scale=0.45]{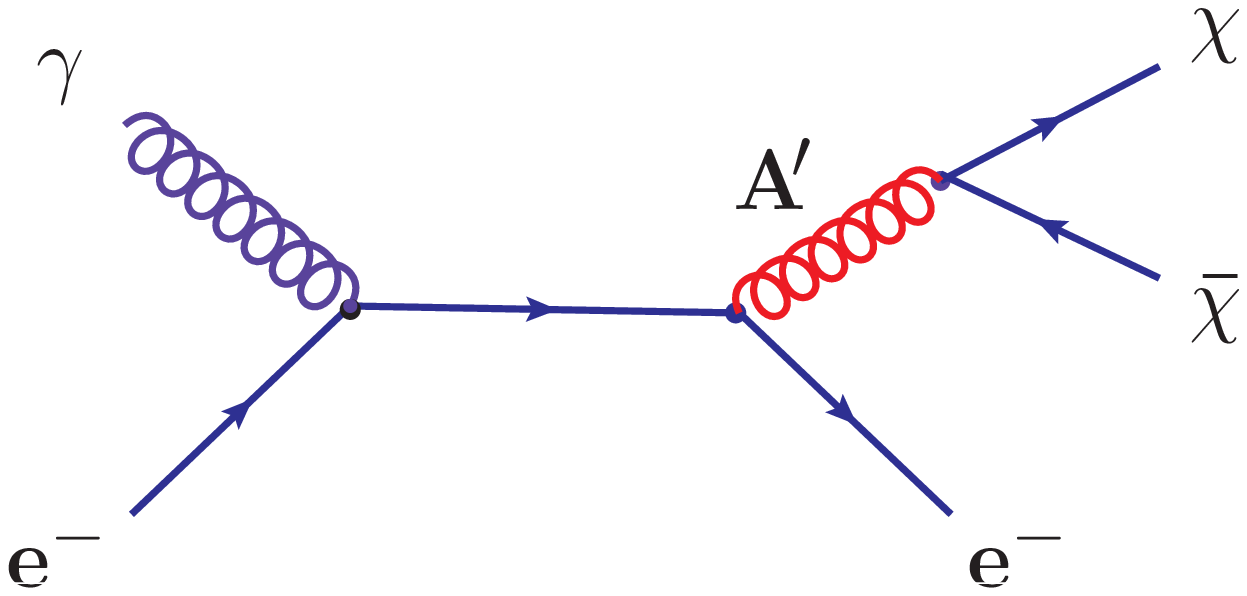}
\includegraphics[scale=0.45]{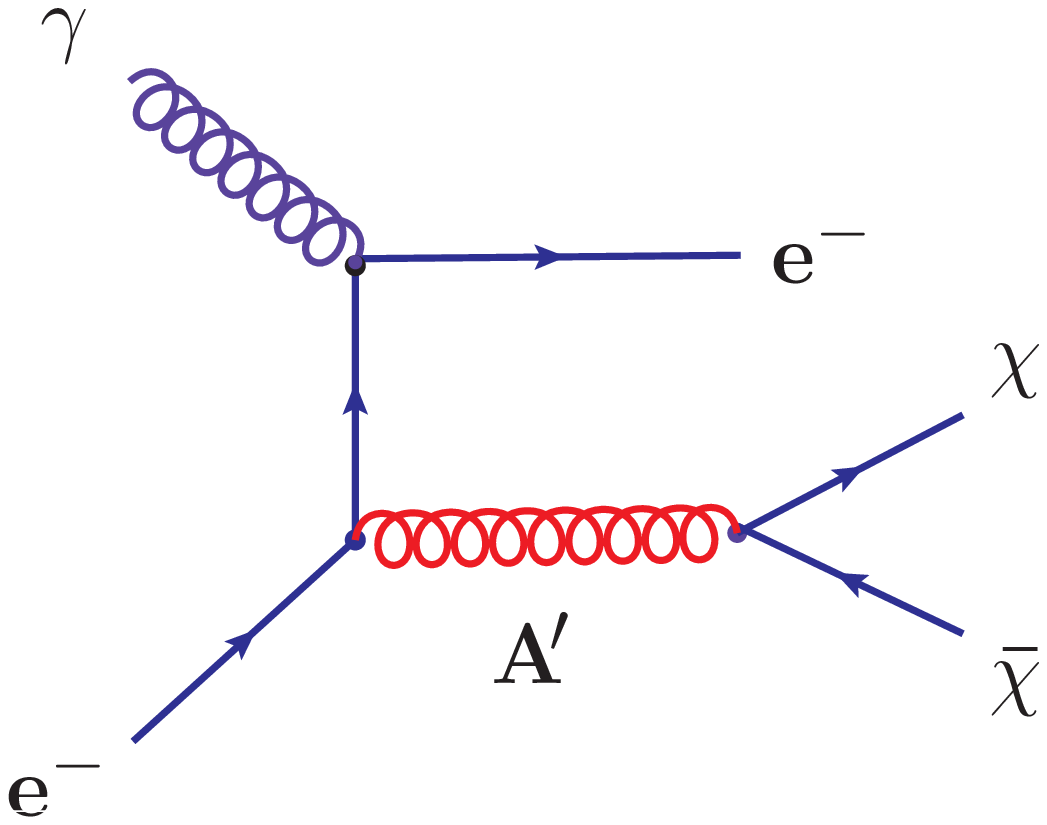} \\
\caption{The diagrams for the production of the dark photon in the
$\gamma e^-$ collision, with its subsequent invisible decay into
dark matter particles.}
\label{fig:DP_prod}
\end{center}
\end{figure}

The main background comes from the SM process
\begin{equation}\label{SM_background_process}
\gamma e^- \rightarrow e^- \nu \bar{\nu} \;,
\end{equation}
see the diagrams in Figs.~\ref{fig:SM_s} and \ref{fig:SM_u}. In
detector it looks like an event with an isolated electron and
missing transverse energy. We apply the cut on the transverse
momenta of the final electron, $p_\perp > 10$ GeV, and  its
rapidity, $|\eta| < 2.5$.  In order to reduce the SM background, we
also impose the cut on an invisible invariant mass, $|m_{A'} -
m_{\mathrm{invis}}| < 5$ GeV. In numerical calculations, especially
in background calculations, CalcHEP program was also used
\cite{CalcHep}. We have used the following statistical significance
($SS$) formula \cite{SS},
\begin{equation}\label{SS_def}
SS = \sqrt{2[(S+B) \,\ln(1 + S/B) - S]} \;,
\end{equation}
where $S$ and $B$ are the numbers of the signal and background
events, respectively. Note that $SS \simeq S/\sqrt{B}$ for $S \ll
B$.
%
%%%%%%%%%%%%%%%%%%%%%%%%%%%%%%%%%%%%%%%%%%%%%%%%%%%%%%%%%%%%%%%%%%%%%%%%%%
% Figure 4. s-channel diagrams for SM process gamma + e -> e + missing E %
%%%%%%%%%%%%%%%%%%%%%%%%%%%%%%%%%%%%%%%%%%%%%%%%%%%%%%%%%%%%%%%%%%%%%%%%%%
\begin{figure}[htb]
\begin{center}
\includegraphics[scale=0.45]{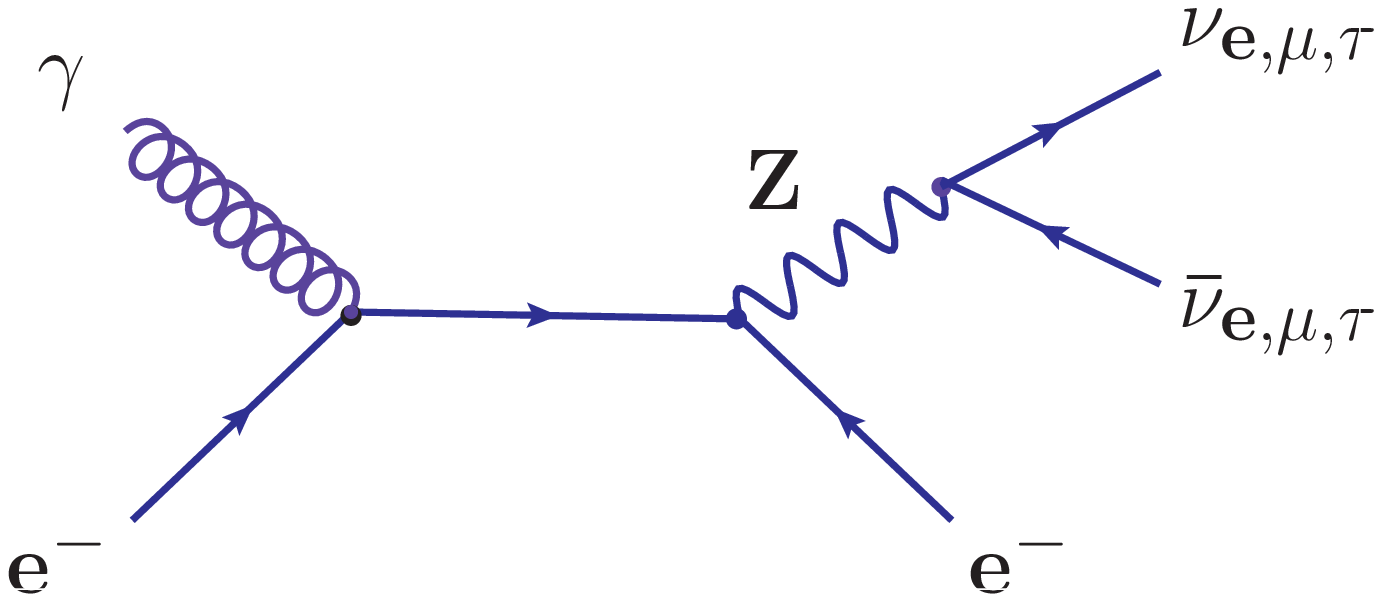}
\includegraphics[scale=0.45]{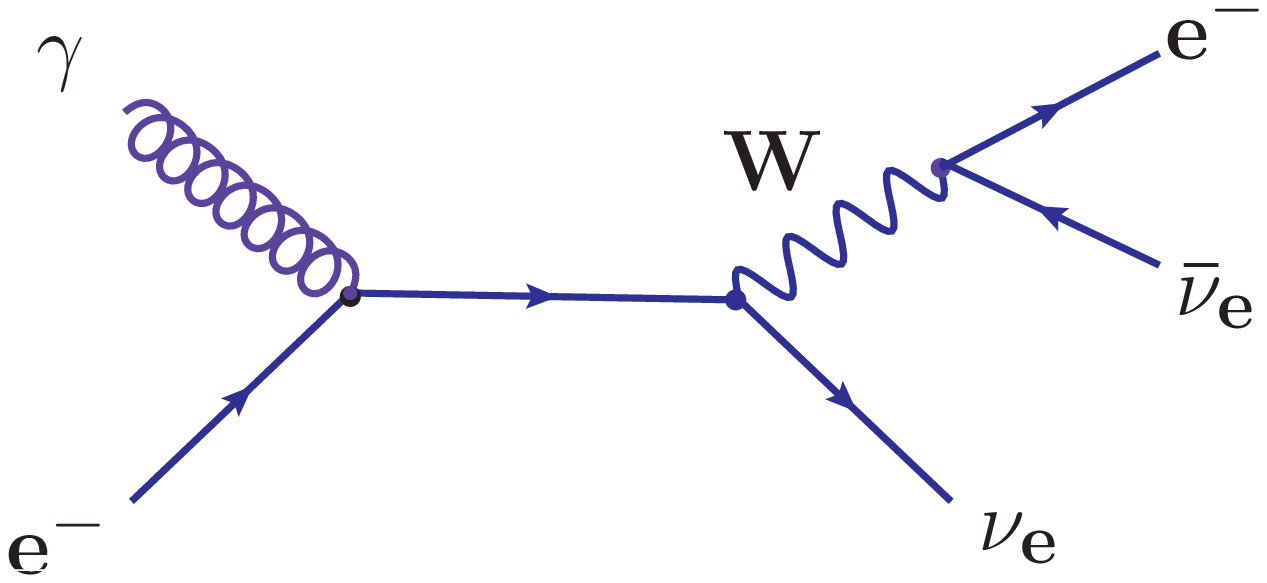}
\caption{The $s$-channel diagrams for the SM process $\gamma e^-
\rightarrow e^- + E{\!\!\!\!/}$.} \label{fig:SM_s}
\end{center}
\end{figure}
%
%%%%%%%%%%%%%%%%%%%%%%%%%%%%%%%%%%%%%%%%%%%%%%%%%%%%%%%%%%%%%%%%%%%%%%%%%%
% Figure 5. t-channel diagrams for SM process gamma + e -> e + missing E %
%%%%%%%%%%%%%%%%%%%%%%%%%%%%%%%%%%%%%%%%%%%%%%%%%%%%%%%%%%%%%%%%%%%%%%%%%%
\begin{figure}[htb]
\begin{center}
\includegraphics[scale=0.45]{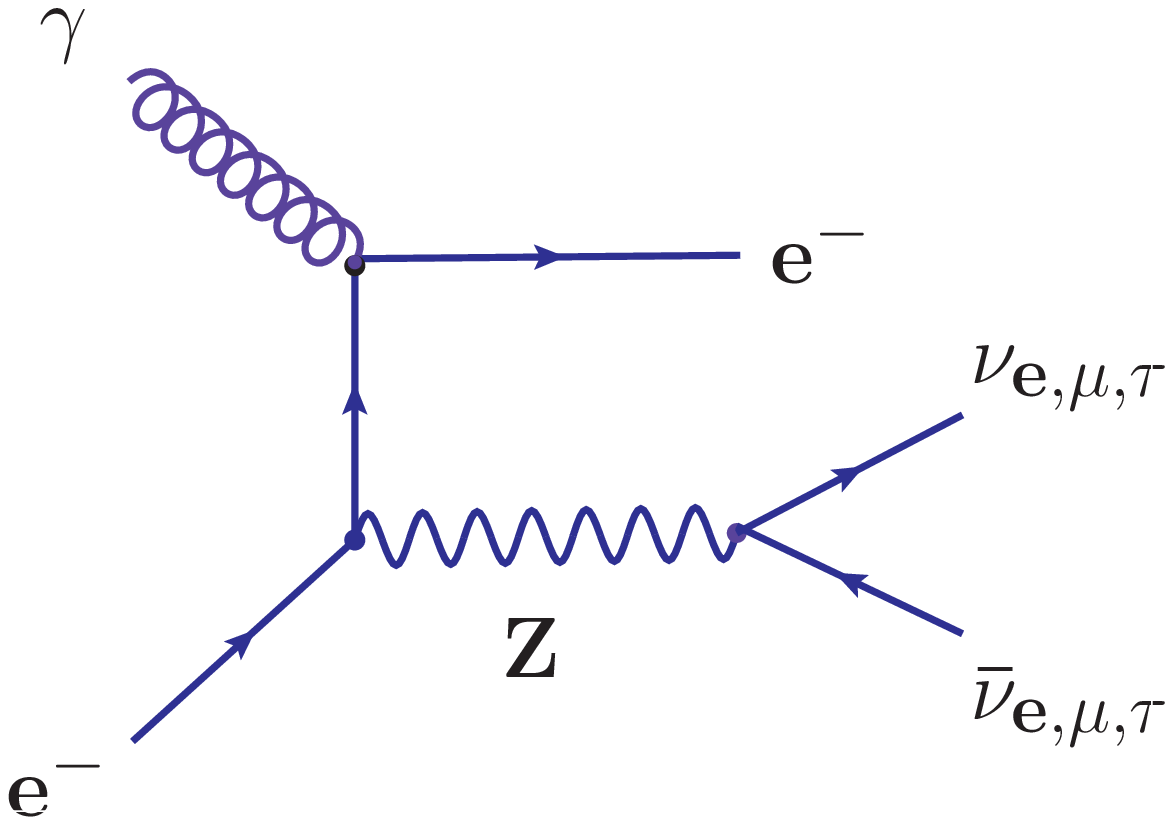}
\includegraphics[scale=0.45]{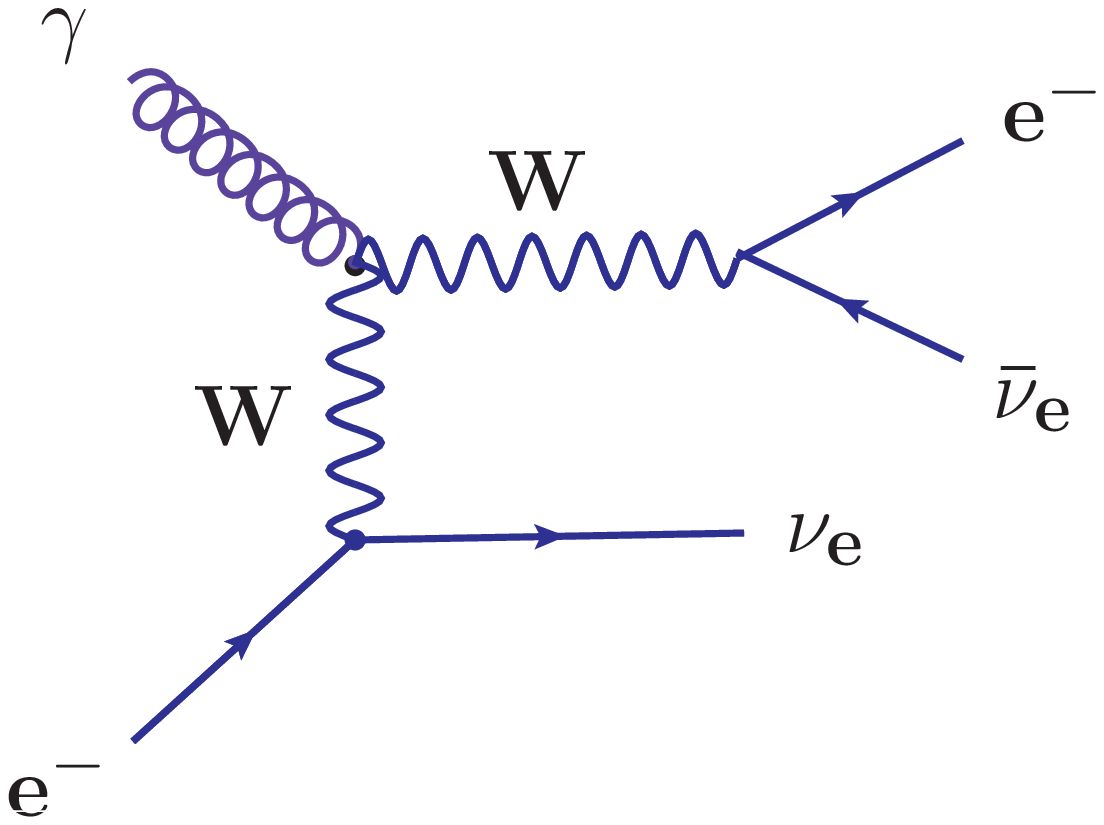}
\includegraphics[scale=0.45]{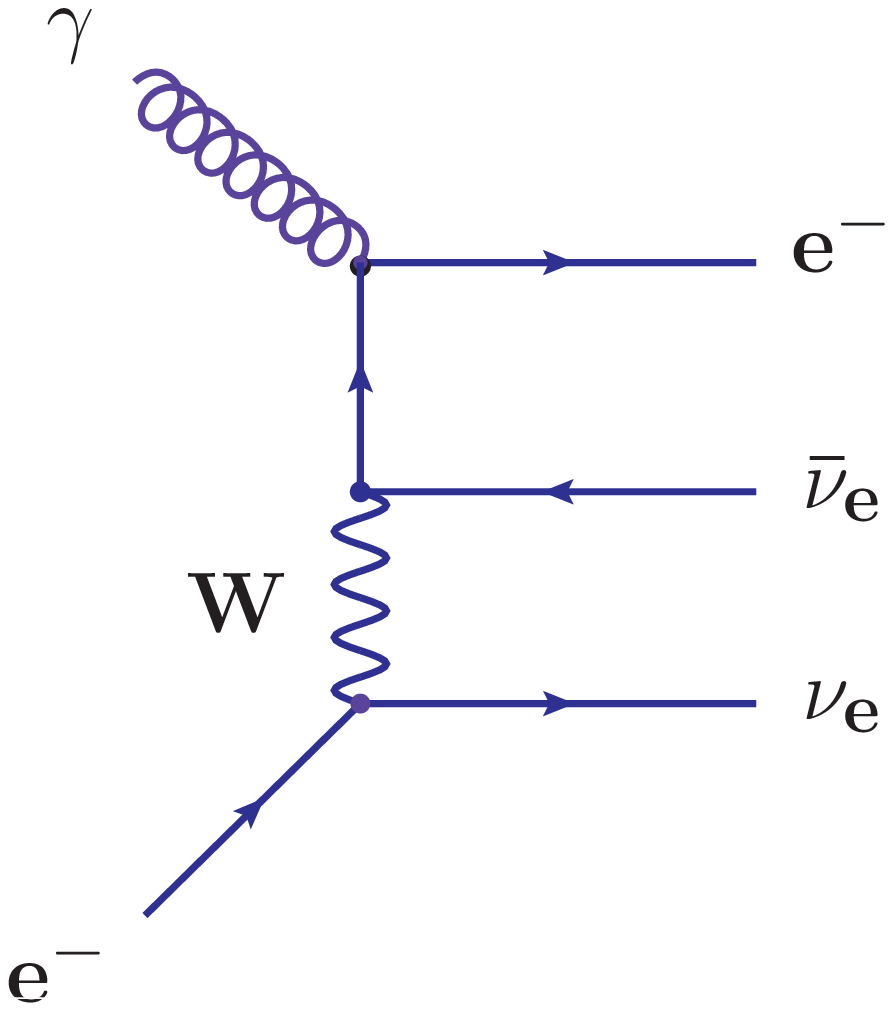}
\caption{The $u$-channel diagrams for the SM process $\gamma e^-
\rightarrow e^- + E{\!\!\!\!/}$.}
\label{fig:SM_u}
\end{center}
\end{figure}

In Figs.~\ref{fig:EDCS_CEPC}-\ref{fig:EDCS_CLIC} we present
differential cross sections of the $\gamma e^- \rightarrow e^- +
E{\!\!\!\!/}$ scattering as functions of the missing energy for a
fixed value of the mixing parameter $\varepsilon$. The results for
the colliders CEPC, ILC, and CLIC are given, for the DP masses
$m_{A'} = 50$ GeV, 100 GeV, and 200 GeV, respectively. In all
figures, red (black) curves correspond to the cross sections with
(without) account of the dark photon contributions. It turns out
that DP is the dominant effect for high missing energy values. We
have also made calculations for different values of $m_{A'}$. Our
calculations show that deviations of the cross sections from the SM
predictions become smaller as the DP mass $m_{A'}$ grows.

%%%%%%%%%%%%%%%%%%%%%%%%%%%%%%%%%%%%%%%%%%%%%%%%%
% Figure 6. Differential cross section for CEPC %
%%%%%%%%%%%%%%%%%%%%%%%%%%%%%%%%%%%%%%%%%%%%%%%%%
\begin{figure}[htb]
\begin{center}
\includegraphics[scale=0.6]{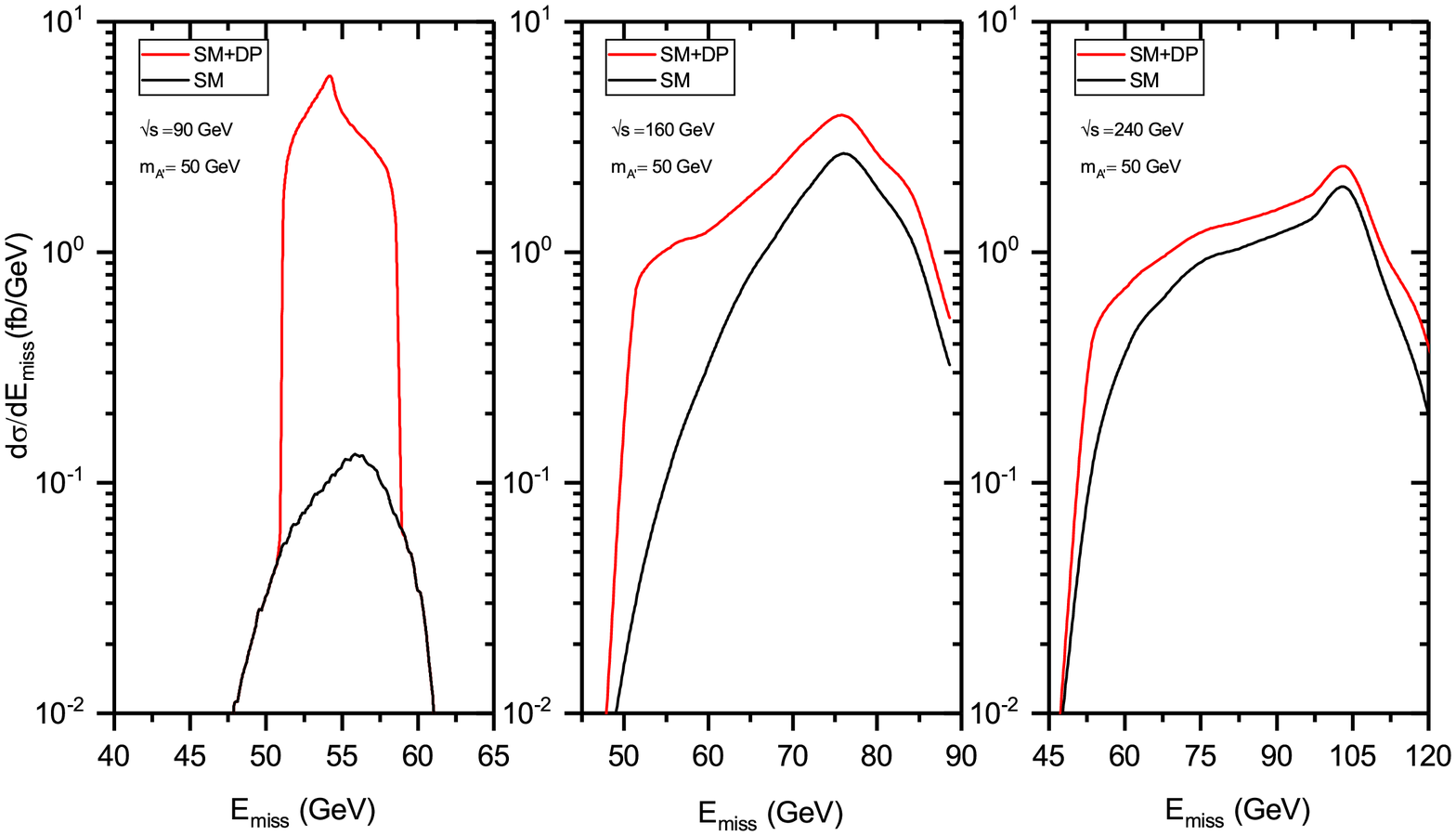}
\caption{The differential cross sections for the unpolarized $\gamma
e^- \rightarrow e^- + E{\!\!\!\!/}$ scattering at the collider CEPC.
The mixing parameter is fixed to be $\varepsilon = 0.1$.}
\label{fig:EDCS_CEPC}
\end{center}
\end{figure}
%
%%%%%%%%%%%%%%%%%%%%%%%%%%%%%%%%%%%%%%%%%%%%%%%%
% Figure 7. Differential cross section for ILC %
%%%%%%%%%%%%%%%%%%%%%%%%%%%%%%%%%%%%%%%%%%%%%%%%
\begin{figure}[htb]
\begin{center}
\includegraphics[scale=0.6]{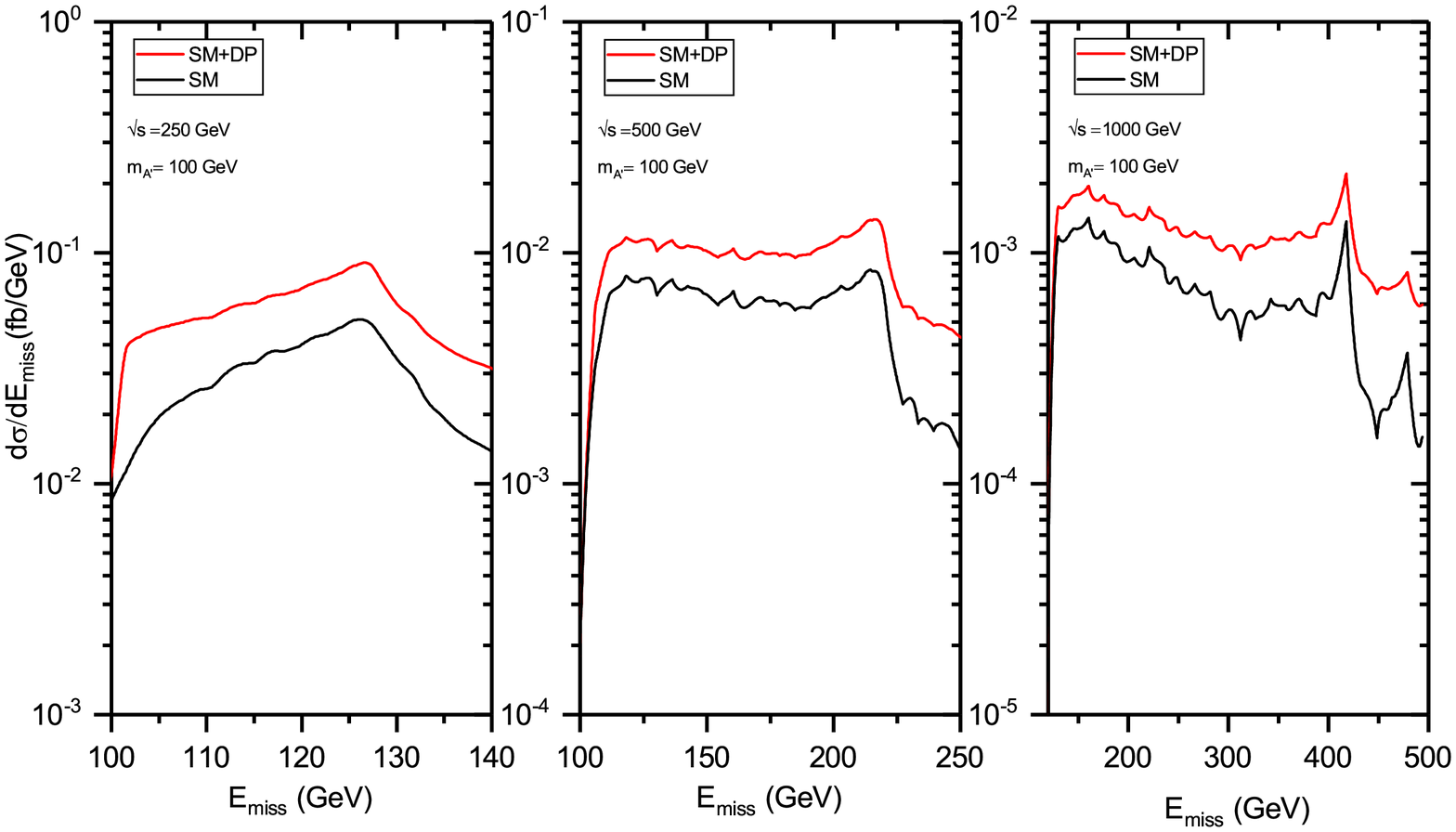}
\caption{The same as in Fig.~\ref{fig:EDCS_CEPC}, but for the
collider ILC.}
\label{fig:EDCS_ILC}
\end{center}
\end{figure}
%
%%%%%%%%%%%%%%%%%%%%%%%%%%%%%%%%%%%%%%%%%%%%%%%%%
% Figure 8. Differential cross section for CLIC %
%%%%%%%%%%%%%%%%%%%%%%%%%%%%%%%%%%%%%%%%%%%%%%%%%
\begin{figure}[htb]
\begin{center}
\includegraphics[scale=0.6]{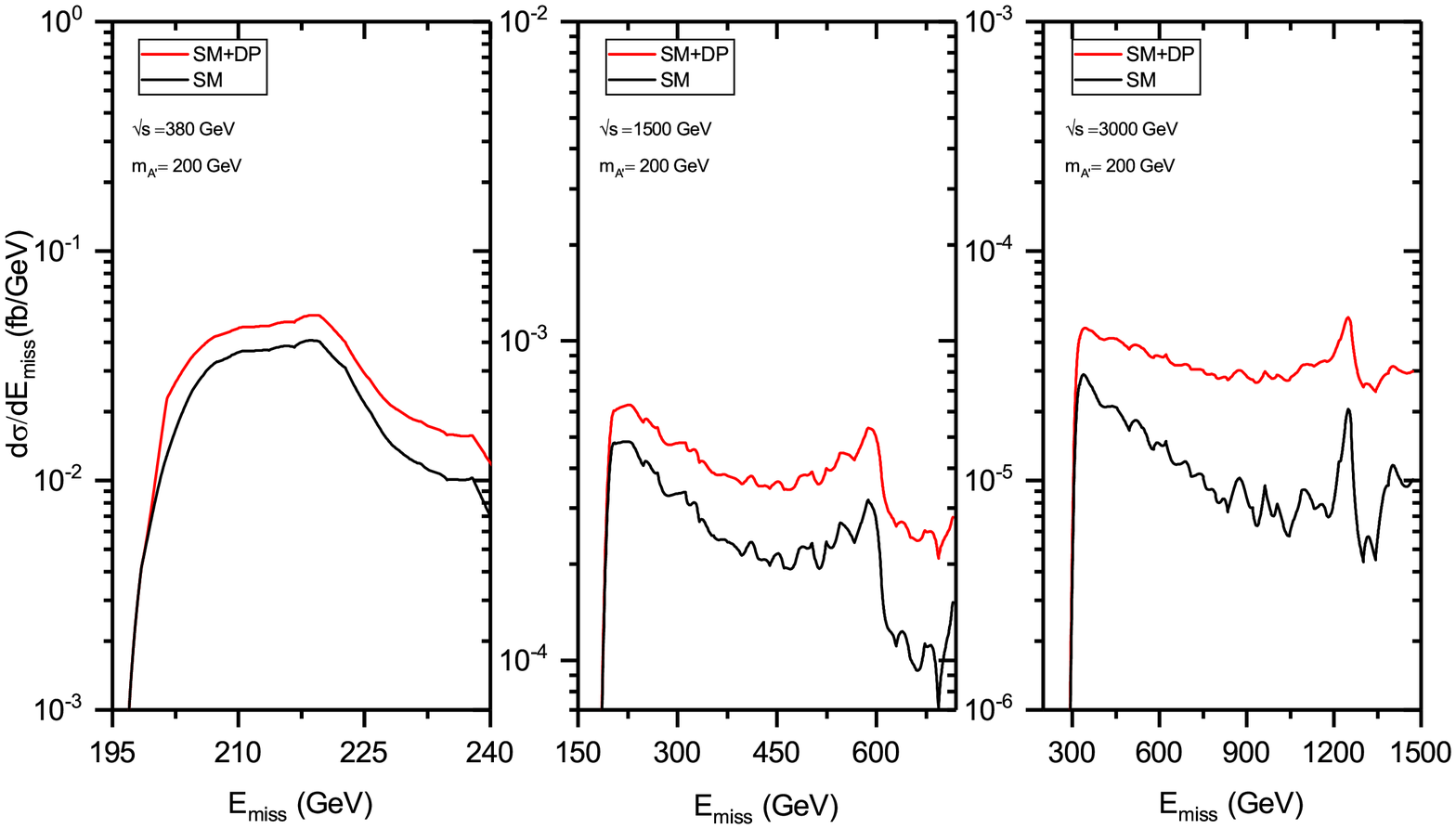}
\caption{The same as in Fig.~\ref{fig:EDCS_CEPC}, but for the
collider CLIC.}
\label{fig:EDCS_CLIC}
\end{center}
\end{figure}

The excluded bounds for the massive DP going to invisible final
states are presented in Figs.~\ref{fig:SSDP}, \ref{fig:PSSDP} in the
plane the kinetic mixing parameter $\varepsilon$ versus DP mass
$m_{A'}$. In Fig.~\ref{fig:SSDP} the results for the unpolarized
$\gamma e^- \rightarrow A' e^-$ collision are given for the CEPC
(left panel), ILC (middle panel), and CLIC (right panel). The
strongest bound on $\varepsilon$ is achieved for $m_{A'} = 10$ GeV
and $\sqrt{s} = 90$ GeV. The bounds for $m_{A'} = 10$ GeV and other
collision energies ($\varepsilon \sim 10^{-3})$ are comparable this
those obtained by the BaBar collaboration, see
Fig.~\ref{fig:current_limits_invis}. The sensitivity decreases as
$m_{A'}$ grows.
%
%%%%%%%%%%%%%%%%%%%%%%%%%%%%%%%%%%%%%%%%%%%%%%%%
% Figure 9. Exclusion bounds. Unpolarized case %
%%%%%%%%%%%%%%%%%%%%%%%%%%%%%%%%%%%%%%%%%%%%%%%%
\begin{figure}[htb]
\begin{center}
\includegraphics[scale=0.6]{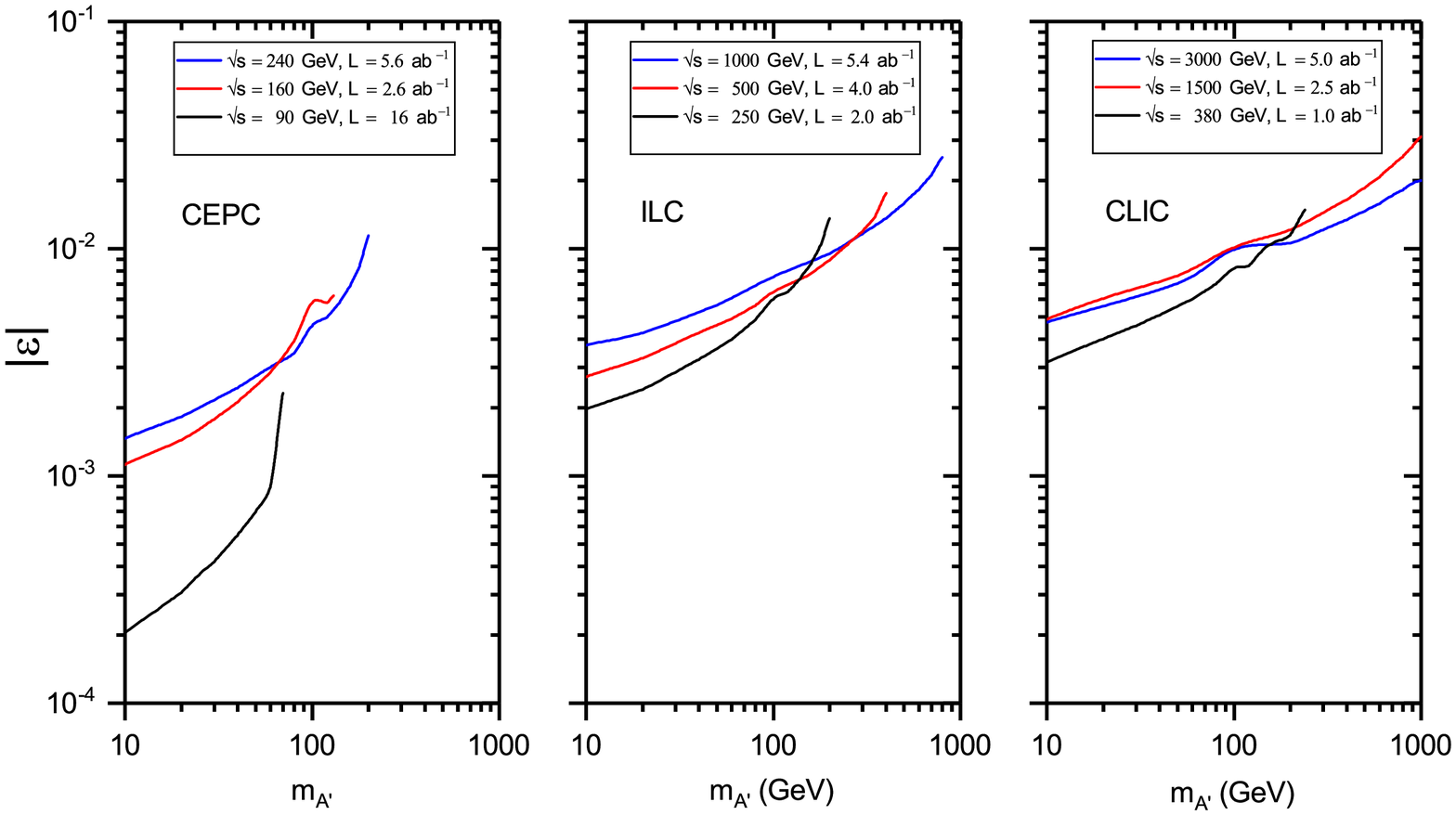}
\caption{The excluded bounds at the 95\% C.L. on the dark photon
mass $m_{A'}$ and kinetic mixing parameter $\varepsilon$ for the
invisible dark photon production in the \emph{unpolarized} $\gamma
e^- \rightarrow A' e^-$ collision.}
\label{fig:SSDP}
\end{center}
\end{figure}

The results for the polarized case are shown in
Fig.~\ref{fig:PSSDP}. The polarized electron sources of the future
linear colliders have been discussed in the current ILC
\cite{ILC_2}, CLIC \cite{CLIC_1}, and CEPC \cite{CEPC_1} designs. We
consider the \emph{unpolarized CB photons} and $\mathrm{P}(e^-) =
80\%$ polarization of the initial electron beam for all three
colliders. By analogy with the ILC project, we assume that for the
polarized $\gamma e^-$ collision, the CEPC integrated luminosity is
50\% less than the integrated luminosity for the unpolarized $\gamma
e^-$ scattering. One can see that the obtained excluded bounds are
on average 25\% better compared to the unpolarized bounds
(Fig.~\ref{fig:SSDP}). For the CLIC, a 10\% improvement takes place
only for the energy $\sqrt{s} = 380$ GeV. As for the electron beam
polarization of $\mathrm{P}(e^-) = -80\%$, our calculations show
that it does not offer any advantage over the unpolarized case. The
reason is that the SM background gets larger for $\mathrm{P}(e^-) =
-80\%$ with respect to the unpolarized collision, while the signal
remains almost the same.
%
%%%%%%%%%%%%%%%%%%%%%%%%%%%%%%%%%%%%%%%%%%%%%%%%%
% Figure 10. Exclusion bounds. Unpolarized case %
%%%%%%%%%%%%%%%%%%%%%%%%%%%%%%%%%%%%%%%%%%%%%%%%%
\begin{figure}[htb]
\begin{center}
\includegraphics[scale=0.6]{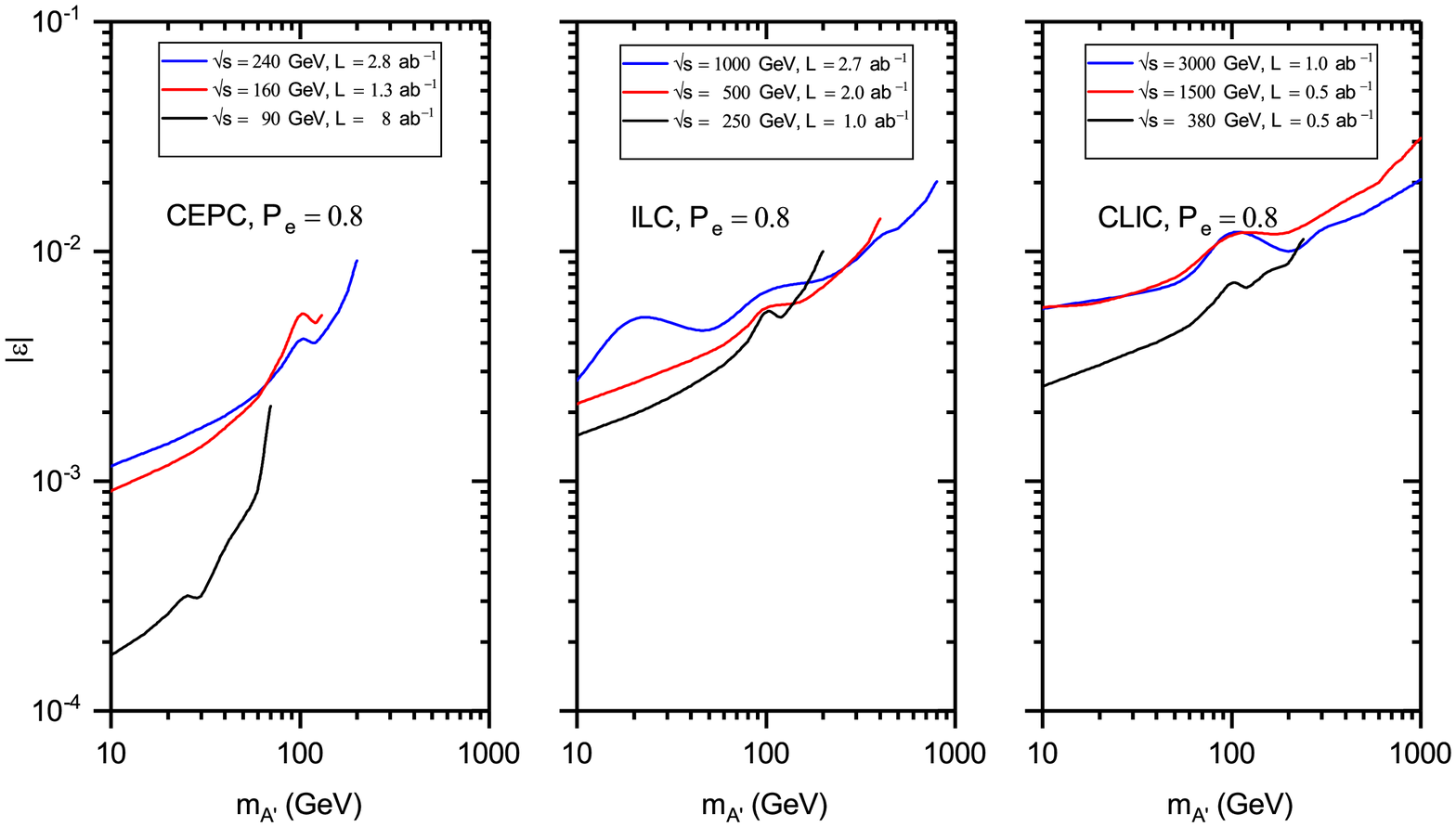}
\caption{The excluded bounds at the 95\% C.L. on the dark photon
mass $m_{A'}$ and kinetic mixing parameter $\varepsilon$ for the
invisible dark photon production in the collision of the
\emph{unpolarized} real photon with the electron whose polarization
is equal to 80\%.}
\label{fig:PSSDP}
\end{center}
\end{figure}

%%%%%%%%%%%%%%%%%%%%%%%
\section{Conclusions} %
%%%%%%%%%%%%%%%%%%%%%%%

In the present paper, we have studied the production of the massive
dark photon (DP) in the $\gamma e^-$ scattering at the future lepton
colliders ILC, CLIC, and CEPC, when the DP decays predominantly into
invisible dark matter particles. The real photons are generated by
the laser Compton backscattering when the soft laser photons collide
with the electron beams. Both the unpolarized and polarized
collisions are studied. For the polarized $\gamma e^- \rightarrow A'
e^-$ process we have assumed that the incoming CB photon is
unpolarized, while the polarization of the electron beam is taken to
be $80\%$ for all three colliders. The region 10 GeV -- 1000 GeV of
the DP mass $m_{A'}$ is considered. The missing energy distributions
for signal and background are presented. We have derived the
excluded regions at the 95\% C.L. in the plane $(\varepsilon,
m_{A'})$, where $\varepsilon$ is the kinetic mixing parameter. Our
excluded bounds for the polarized collisions at the ILC and CEPC are
approximately 25\% stronger compared to the unpolarized excluded
bounds. Note that up to now the production of the DP at future
lepton colliders (both for the visible and invisible DP decays) was
studied only for the $e^+e^-$ mode of these colliders.

%%%%%%%%%%%%%%%%%%%%%%%%%%%%%%%%%%%%%%%%%%%%%%%%%%%%%%%%%%%%%%%%%%%%%

%%%%%%%%%%%%%%
% Appendix A %
%%%%%%%%%%%%%%

\setcounter{equation}{0}
\renewcommand{\theequation}{A.\arabic{equation}}

\section*{Appendix A}
\label{app:A}

The matrix element of the process \eqref{process} is given by the
sum of $s$- and $u$-channel diagrams presented in
Fig.~\ref{fig:DP_prod}. Correspondingly, we get
\begin{equation}\label{M2}
|F(s, \cos\theta)|^2 = (\varepsilon e^2)^2 \,\big[ |F_s|^2 + |F_u|^2
+ (F_s F_u^* + F_s^* F_u) \big] \;,
\end{equation}
where
\begin{align}\label{Ms_Mu_Mint}
|F_s|^2 &= \frac{2}{(s - m^2)^2} \left[ -su + m^2(3s + u +
2m_{A'}^2) + m^4 \right] ,
\nonumber \\
|F_u|^2 &= \frac{2}{(u - m^2)^2} \left[ -su + m^2(3u + s +
2m_{A'}^2) + m^4 \right] \;,
\nonumber \\
F_s F_u^* &= F_s^* F_u = \frac{2}{(s - m^2)(u - m^2)}
\nonumber \\
&\times \left[ m_{A'}^2(s + u) - m_{A'}^4 + m^2(s + u - 2m_{A'}^2) +
2m^4 \right] .
\end{align}
The Mandelstam variable $u$ is equal to
\begin{align}\label{u}
u &= m_{A'}^2 + m^2 - \frac{1}{2s}\Big[ (s + m^2)(s + m_{A'}^2 -
m^2)
\nonumber \\
&+ (s - m^2)\sqrt{(s + m_{A'}^2 - m^2)^2 - 4s \,m_{A'}^2}\cos\theta
\Big] ,
\end{align}
where $\theta$ is a scattering angle of the DP in the center-of-mass
frame of the colliding particles (see Fig.~\ref{fig:DP_process}).
For $m_{A'}=0$, $\varepsilon = 1$, the above formulas coincide with
well-known formulas for the Compton scattering \cite{Compton_scatt}.

%%%%%%%%%%%%%%%%%%%%%%%%%%%%%%%%%%%%%%%%%%%%%%%%%%%%%%%%%%%%%%%%%%%%%

%%%%%%%%%%%%%%
% References %
%%%%%%%%%%%%%%

%%%%%%%%%%%%%%%%%%%%%%%%%%%%%%%%%%%%%%%%%%%%%%%%%%%%%%%%%%%%%%%%%%%%%

%%%%%%%%%%%%%%%
\end{document}